\documentclass[a4paper]{article}

\usepackage{kotex}
\usepackage{amsmath}
\usepackage{amssymb}
\usepackage{hhline}
\usepackage{amsthm}
\usepackage{commath}
\usepackage{mathtools}
\usepackage{nicefrac}
\usepackage{xfrac}
\usepackage{multirow}
\usepackage{INTERSPEECH2020}
\usepackage{color}
\usepackage{hyperref}
\DeclareMathOperator*{\argmax}{arg\,max}

\usepackage{pifont}
\newcommand{\cmark}{\ding{51}}%
\newcommand{\xmark}{\ding{55}}%
\usepackage{enumitem}

\title{Phase-aware Single-stage Speech Denoising and Dereverberation with U-Net}
\name{Hyeong-Seok Choi$^{1,2}$, Hoon Heo$^2$, Jie Hwan Lee$^{1,2}$, Kyogu Lee$^{1,2}$}
\address{
  $^1$Music and Audio Research Group (MARG), Seoul National University\\
  $^2$Supertone Inc.}
\email{\{kekepa15, wiswisbus, kglee\}@snu.ac.kr, hoon@supertone.ai}

\begin{document}

\maketitle
\begin{abstract}
  In this work, we tackle a denoising and dereverberation problem with a single-stage framework.
  Although denoising and dereverberation may be considered two separate challenging tasks, and thus, two modules are typically required for each task, we show that a single deep network can be shared to solve the two problems.
  To this end, we propose a new masking method called phase-aware $\beta$-sigmoid mask (PHM), which reuses the estimated magnitude values to estimate the clean phase by respecting the triangle inequality in the complex domain between three signal components such as mixture, source and the rest. 
  Two PHMs are used to deal with direct and reverberant source, which allows controlling the proportion of reverberation in the enhanced speech at inference time.
  In addition, to improve the speech enhancement performance, we propose a new time-domain loss function and show a reasonable performance gain compared to MSE loss in the complex domain.
  Finally, to achieve a real-time inference, an optimization strategy for U-Net is proposed which significantly reduces the computational overhead up to 88.9\% compared to the naïve version\footnote{Audio samples link: \href{https://tinyurl.com/ycndlmfm}{https://tinyurl.com/ycndlmfm}}.

\end{abstract}

\section{Introduction}
Corrupted speech signal by noise and reverberation is one of the most common signals we hear in our everyday life.
The desire to listen to clean speech signal, therefore, is strong let alone its usage for machine such as speech recognition system.
While there has been numerous studies to address the problem of single channel denoising and dereverberation, only few have tried to solve this problem with a single deep learning model \cite{sun2018enhanced}.
This motivates us to tackle this real-world problem using a single-stage deep learning model.
We address this problem by dissecting the elements that compose the noisy-reverberant mixture, that is, noise, direct source, and reverberation.
By dissecting each part of the mixture, one could handle each element at hand and even mix them with a desired proportion.
Note that this is a desirable property for users as the effective amount of reverberation is important to achieve better speech intelligibility for both impaired and nonimpaired listeners \cite{bradley2003importance, hu2014effects}.

The key contributions of our proposed approach are three folds.
First, to suppress the noise and reverberation, we propose a new type of complex-valued mask called phase-aware $\beta$-sigmoid mask (PHM).
While the complex-valued mask suggested by \cite{williamson2016complex} separately estimates the real and imaginary part of a complex spectrogram, we believe the phase part of it can be effectively estimated by reusing an estimated magnitude value of it in a trigonometric perspective as suggested in \cite{wang2019masking}.
The major difference between PHM and the suggested approach in \cite{wang2019masking} is that PHM is designed to respect the triangular relationship between mixture, source and the rest, and hence the sum of the estimated source and the rest is always equal to the mixture.
By exploiting this property, we train the deep network to output two different PHMs simultaneously to effectively deal with both denoising and derverbration problem.
Second, we propose a new time-domain loss function, an emphasized multi-scale cos similarity loss function.
A time-domain loss function has recently been used as a popular loss function \cite{choi2019phase, yao2019coarse, wang2018end, koizumi2020speech, le2019sdr}.
To better design the time-domain cos similarity loss function proposed in \cite{choi2019phase}, we change it into a multi-scale version of it with proper emphasis functions and show the effectiveness of it.
Finally, we suggest an optimization strategy for two-dimensional U-Net to significantly reduce the computational inefficiency in runtime.

\section{Related Works}
Recently, there has been an increasing interests in phase-aware speech enhancement because of the sub-optimality of reusing the phase of mixture signal.
The first work that tried to address this problem was by using phase-sensitive mask (PSM) \cite{erdogan2015phase}. PSM estimates the real-part of the signal which is still sub-optimal.
As a more direct remedy for this, a complex masking \cite{williamson2016complex, choi2019phase, wang2019masking} or complex spectral mapping \cite{tan2019complex} has also been proposed to estimate a clean phase part.
Another line of research is to sequentially estimate the clean phase part using an additional sub-module \cite{takahashi2018phasenet, afouras2018conversation, yin2019phasen}. This, however, is limited in that it requires an additional module resulting in inefficient computation.
While most of these works tried to estimate the clean phase by using phase mask or an additional network, the absolute phase difference between mixture and source can be actually computed using the law of cosines using the estimated magnitude values as the three sides of a triangle \cite{mowlaee2012phase, mowlaee2014time}.
Inspired by this, \cite{wang2019deep} proposed to estimate a rotational direction of the absolute phase difference using a sign-prediction network.

The efforts to deal with denoising or dereverberation using deep networks have been tried in many works. Recently, \cite{zhao2018two, maciejewski2019whamr} tried to address this problem with two modules for each task. 
We believe, however, a two-stage framework is not necessary and can be achieved using a single deep network.

\section{Single-stage Denoising and Dereverberation}
A noisy-reverberant mixture signal $\bm{x}$ is commonly modeled as the sum of additive noise $\bm{y}^{(n)}$ and reverberant source $\tilde{\bm{y}}$, where $\tilde{\bm{y}}$ is a result of convolution between room impulse response (RIR) $\bm{h}$ and dry source $\bm{y}$ as follows,
$\bm{x} = \tilde{\bm{y}} + \bm{y}^{(n)} = \bm{h} \circledast \bm{y} + \bm{y}^{(n)}$.
More concretely, we can break down $\bm{h}$ into two parts, that is, direct path part $\bm{h}^{(d)}$ which does not includes the reflection path and the rest of the part $\bm{h}^{(r)}$ that includes all the reflection paths as follows, 
$\bm{x} = (\bm{h}^{(d)} + \bm{h}^{(r)}) \circledast \bm{y} + \bm{y}^{(n)} = \bm{h}^{(d)} \circledast \bm{y} + \bm{h}^{(r)} \circledast \bm{y} + \bm{y}^{(n)} = \bm{y}^{(d)} + \bm{y}^{(r)} + \bm{y}^{(n)}$,
where $\bm{y}^{(d)}$ and $\bm{y}^{(r)}$ denotes direct path source and reverberation, respectively.
In this setting, our goal is to separate $\bm{x}$ into three elements $\bm{y}^{(d)}$, $\bm{y}^{(r)}$, and $\bm{y}^{(n)}$.
Each of the corresponding time-frequency $(t,f)$ representations computed by STFT is denoted as $X_{t,f} \in \mathbb{C}$, $Y^{(d)}_{t,f}\in \mathbb{C}$, $Y^{(r)}_{t,f} \in \mathbb{C}$, $Y^{(n)}_{t,f} \in \mathbb{C}$, and the estimated values will be denoted by the hat operator $\hat{\, \cdot \,}$. 

\subsection{Phase-aware \texorpdfstring{$\beta$}{}-sigmoid mask}
Designing a mask that is not limited to output the optimal value of ideal mask requires two conditions to satisfy.
First, the range of magnitude mask should not be limited.
Second, the mask has to be complex-valued so that it can correct both the magnitude part and phase part of the mixture signal.
The proposed phase-aware $\beta$-sigmoid mask (PHM) is designed to handle both conditions while systemically restricting the sum of estimated complex values to be exactly the value of mixture, $X_{t,f} = Y^{(k)}_{t,f} + Y^{(\lnot k)}_{t,f}$.
The PHM separates the mixture $X_{t,f}$ in STFT domain into two parts as \textit{one-vs-rest} approach, that is, the signal $Y^{(k)}_{t,f}$ and the sum of the rest of the signals $Y^{(\lnot k)}_{t,f} = X_{t,f}-Y^{(k)}_{t,f}$, where index $k$ could be one of direct path source ($d$), reverberation ($r$), and noise ($n$) in our setting, $k \in \{d, r, n\}$.
The complex-valued mask $M^{(k)}_{t,f} \in \mathbb{C}$ estimates the magnitude and phase value of the source of interest $k$.
The mask is composed of two parts, (1) magnitude mask estimation, (2) phase estimation by reusing the magnitude estimation from (1) and two-class sign prediction.

First, the network outputs the magnitude part of two masks $\abs{M^{(k)}_{t,f}}$ and $\abs{M^{(\lnot k)}_{t,f}}$ with sigmoid function $\sigma^{(k)}(\bm{z}_{t,f})$ multiplied by coefficient $\beta_{t,f}$ as follows,
\begin{equation}
\label{eq:magnitude_mask}
\abs{M^{(k)}_{t,f}} = \beta_{t,f} \cdot \sigma^{(k)}(\bm{z}_{t,f}) = \beta_{t,f} \cdot \frac{1}{1+e^{-(z^{(k)}_{t,f} - z^{(\lnot k)}_{t,f})}}
\end{equation}
where $z^{(k)}_{t,f}$ is the output located at $(t,f)$ from the last layer of neural-network function $\psi^{(k)}(\phi)$, and $\phi$ is a function composed of network layers before the last layer.
$\abs{M^{(k)}_{t,f}}$ serves as a magnitude mask to estimate source $k$ and the value of it ranges from 0 to $\beta_{t,f}$.
The role of $\beta_{t,f}$ is to design a mask that is close to the optimal mask with a flexible magnitude range so that the value is not bounded between 0 and 1 unlike the typically used sigmoid mask.
In addition, because the sum of the complex valued masks $M^{(k)}_{t,f}$ and $M^{(\lnot k)}_{t,f}$ must compose a triangle, it is reasonable to design a mask that satisfies the triangle inequalities, that is, $\abs{M^{(k)}_{t,f}} + \abs{M^{(\lnot k)}_{t,f}}$ $\geq 1$  and $\abs{\, \abs{M^{(k)}_{t,f}} - \abs{M^{(\lnot k)}_{t,f}}} \leq 1$.
To address the first inequality we designed the network to output $\beta_{t,f}$ from the last layer with a softplus activation function as follows, $\beta_{t,f} = 1+ \texttt{softplus}((\psi_{\beta}(\phi))_{t,f})$, where $\psi_{\beta}$ denotes an additional network layer to output $\beta_{t,f}$. The second inequality can be satisfied by clipping the upper bound of the $\beta_{t,f}$ by $1 \mathbin{/} \abs{\, \sigma^{(k)}(\bm{z}_{t,f}) - \sigma^{(\lnot k)}(\bm{z}_{t,f})}$.

Once the magnitude masks are decided, we can construct a phase mask $e^{j\theta_{t,f}^{(k)}}$. 
Given the magnitudes as three sides of a triangle, we can compute the cosine of absolute phase difference $\Delta \theta_{t,f}^{(k)}$ between the mixture and source $k$ as follows,
$\cos(\Delta \theta_{t,f}^{(k)}) = \nicefrac{ (1+\abs{M_{t,f}^{(k)}}^2 - \abs{M_{t,f}^{(\lnot k)}}^2 ) } {(2 \, \abs{M_{t,f}^{(k)}})}$.
Next, the rotational direction (clockwise or counterclockwise) for phase correction can be decided by estimating sign value $\xi_{t,f} \in \{-1, 1\}$ as follows,
\begin{equation}
\label{eq:phase_mask}
e^{j\theta_{t,f}^{(k)}} = \cos(\Delta \theta_{t,f}^{(k)}) + j \xi_{t,f}\sin(\Delta \theta_{t,f}^{(k)}).
\end{equation}
Two-class straight-through Gumbel-softmax estimator was used to estimate $\xi_{t,f}$ \cite{jang2016categorical}. It allows us to discretize the output of the Gumbel-softmax function $\gamma^{(i)}$ with $\argmax$ and still be able to train the network in an end-to-end manner using a continuous approximation in the backward pass.
$\xi_{t,f}$ is defined as follows,
\begin{equation}
\label{eq:sign}
\xi_{t,f} = 
    \begin{cases}
        -1, & \gamma^{(0)}(\bm{q}_{t,f}) > \gamma^{(1)}(\bm{q}_{t,f})\\
        1, & \text{otherwise}
    \end{cases}
\end{equation}
where $\gamma^{(i)}(\bm{q}_{t,f})$ is defined as follows,
\begin{equation}
\label{eq:gumbel}
\gamma^{(i)}(\bm{q}_{t,f})=\frac{e^{q^{(i)}_{t,f}}}{\sum_{i}{e^{q^{(i)}_{t,f}}}} = \frac{e^{((\psi_{i}(\phi))_{t,f} + g_i)\mathbin{/}\tau}}{\sum_{i}{e^{((\psi_{i}(\phi))_{t,f} + g_i)\mathbin{/}\tau}}},
\end{equation}
and $g_0$ and $g_1$ are samples from Gumbel$\left(0, 1\right)$, $\psi_i$ is an additional network layer to output logit value $q^{(i)}_{t,f}$, and $\tau$ is a temperature parameter for Gumbel-softmax.
Finally, $M^{(k)}_{t,f}$ is defined as follows,
\begin{equation}
M^{(k)}_{t,f} = \abs{M^{(k)}_{t,f}}e^{j\theta_{t,f}^{(k)}}.
\end{equation}

\subsection{Masking from the perspective of quadrangle}
\begin{figure}[htbp]
\centering
\includegraphics[scale=0.4]{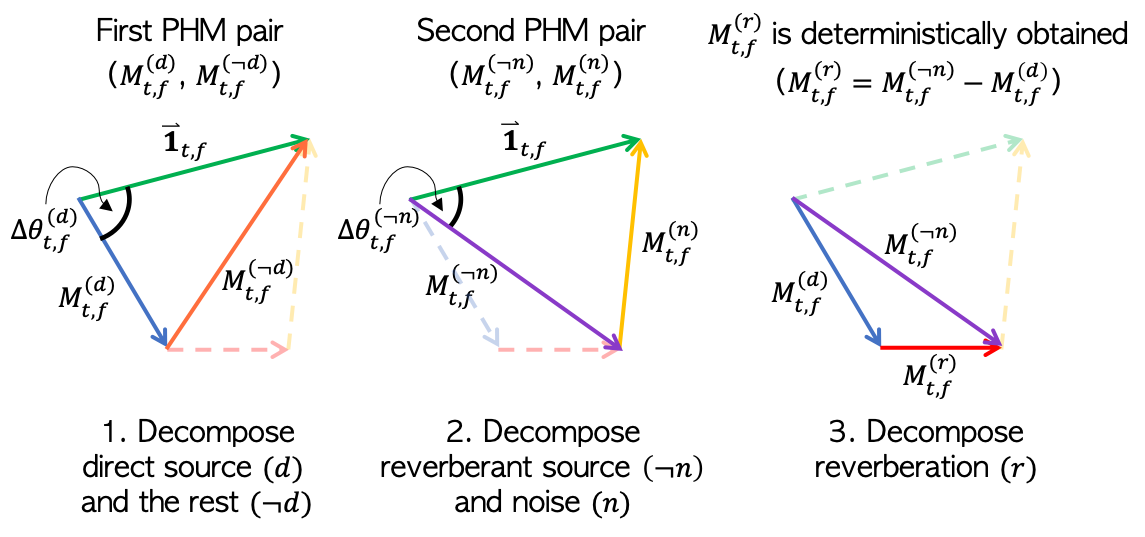}
\caption{The illustration of masks on quadrangle}
\label{fig:quadrangle}
\end{figure}
As we desire to extract both direct source and reverberant source, two pairs of PHMs are used for each of them.
The first pair of masks separates direct source and the rest of the component, denoted as $M^{(d)}_{t,f}$ and $M^{(\lnot d)}_{t,f}$.
The second pair of masks separates noise and reverberant source component, denoted as $M^{(n)}_{t,f}$ and $M^{(\lnot n)}_{t,f}$.
Since PHM guarantees the mixture and separated components to construct a triangle in the complex STFT domain, the outcome of the separation can be seen from the perspective of quadrangle as in Fig \ref{fig:quadrangle}.
In this setting, as the three sides and two side angles are already determined by the two pairs of PHMs, the last fourth side of quadrangle, the reverberation component $\hat{Y}^{(r)}_{t,f}$, is uniquely decided.

\subsection{Emphasized multi-scale cosine similarity loss}
Learning to maximize cosine similarity can be regarded as maximizing the signal-to-distortion ratio (SDR) \cite{choi2019phase}. Cosine similarity loss $C$ between estimated signal $\hat{\bm{y}}^{(k)} \in \mathbb{R}^{N} $ and ground truth signal $\bm{y}^{(k)} \in \mathbb{R}^{N}$  is defined as follows,
\begin{equation}
\label{eq:cossim}
	C(\bm{y}^{(k)},\hat{\bm{y}}^{(k)}) = -\frac{<\bm{y}^{(k)},\hat{\bm{y}}^{(k)}>}{\norm{\bm{y}^{(k)}}  \norm{\hat{\bm{y}}^{(k)}}},
\end{equation}
where $N$ denotes the temporal dimensionality of a signal and $k$ denotes the type of signal ($k \in \{d,r,n\}$).
Consider a sliced signal $\bm{y}^{(k)}_{[\frac{N}{M}(i-1):\frac{N}{M}i]}$, where $i$ denotes the segment index and $M$ denotes the number of segment.
By slicing the signal and normalize it by its norm, each sliced segment is considered as an unit for computing $C$.
 Therefore, we hypothesize that it is important to choose a proper segment length unit $\frac{N}{M}$ when computing $C$.
 In our case, we used multiple settings of segment lengths $g_{j}=\frac{N}{M_{j}}$ as follows,
\begin{equation}
\label{eq:multiscale}
	\mathcal{L}(\bm{y}^{(k)},\hat{\bm{y}}^{(k)}) = \sum_{j} \frac{1}{M_j}\sum_{i=1}^{M_j} C(\bm{y}^{(k)}_{[g_{j}(i-1):g_{j}i]},\hat{\bm{y}}^{(k)}_{[g_{j}(i-1):g_{j}i]}),
\end{equation}
where $M_{j}$ denotes the number of sliced segments. 
In our case the set of $g_j$\textquotesingle s was chosen as follows, $g_j \in \{4064, 2032, 1016, 508\}$, assuming they moderately cover the range of duration of phonemes in speech.

To further improve the design of the loss function, we applied two simple techniques --- 1. pre-emphasis ($\pi$) and 2. $\mu$-law encoding ($\mu$) --- on signals. 
As most of the speech signal components are concentrated in the lower frequency bands, we found that applying pre-emphasis on loss function can help penalize high frequency components.
In addition, since the samples of speech signals are usually centered around zero, we found that it is helpful to use 16-bit $\mu$-law encoding as it distributes samples more uniformly by the nature of continuous logarithmic transform.
The proposed loss function $\mathcal{L}^{+}$ is defined as follows,
\begin{equation}
\begin{split}
\label{eq:emphasized}
	\mathcal{L}^{+}(\bm{y}^{(k)},\hat{\bm{y}}^{(k)}) &= \mathcal{L}(\bm{y}^{(k)},\hat{\bm{y}}^{(k)}) + \mathcal{L}(\pi(\bm{y}^{(k)}),\pi(\hat{\bm{y}}^{(k)})) \\
	& + \mathcal{L}(\mu(\pi(\bm{y}^{(k)})),\mu(\pi(\hat{\bm{y}}^{(k)})))
\end{split}
\end{equation}
Finally, we used the proposed loss function for every $k$ and $\lnot k$ combinations as follows, 
\begin{equation}
 \mathcal{L}_\text{final} = \sum_{k} ( \mathcal{L}^{+}(\bm{y}^{(k)},\hat{\bm{y}}^{(k)}) + \mathcal{L}^{+}(\bm{y}^{(\lnot k)},\hat{\bm{y}}^{(\lnot k)}) ).
\end{equation}

\section{Optimization for Real-Time U-Net}
\label{section:optimization}
To connect each encoder layer with its corresponding decoder, U-Net is often composed of convolutional layers with zero padding for dynamic input sizes. Without zero padding, the valid size of input and output is uniquely determined by the kernel sizes and strides. This obviously takes less computation and allows to keep only the essential part. In our real-time setting, the input spectrogram has 253 frequency bins and 65 frames by discarding the four lowest bins from the original spectrogram with a 512-point FFT, assuming that 16 kHz speech signals have no significant spectral component below 93.75 Hz. 
We followed the network architecture of the real-valued U-Net proposed in \cite{choi2019phase} (\texttt{model10} and \texttt{model20} specifically) with the modification of the last layer to output PHM. All the batch normalizations were fused into convolution filters.

In the encoder, the naïve implementation of U-Net repeatedly performs the same computation that has already been computed previously.
This redundancy can be efficiently reduced by caching the pre-computed values using queues.
Likewise, we utilized a similar concept with 2D convolution, but more than one queues are needed for the strided convolution in each layer.
The number of required queues for depth $d$ is derived by $\prod_{l=1}^{d}{s_l}$, where $s_l$ denotes the temporal stride of the $l$-th encoder layer.

\begin{figure}[!t]
\centering
\includegraphics[scale=0.47]{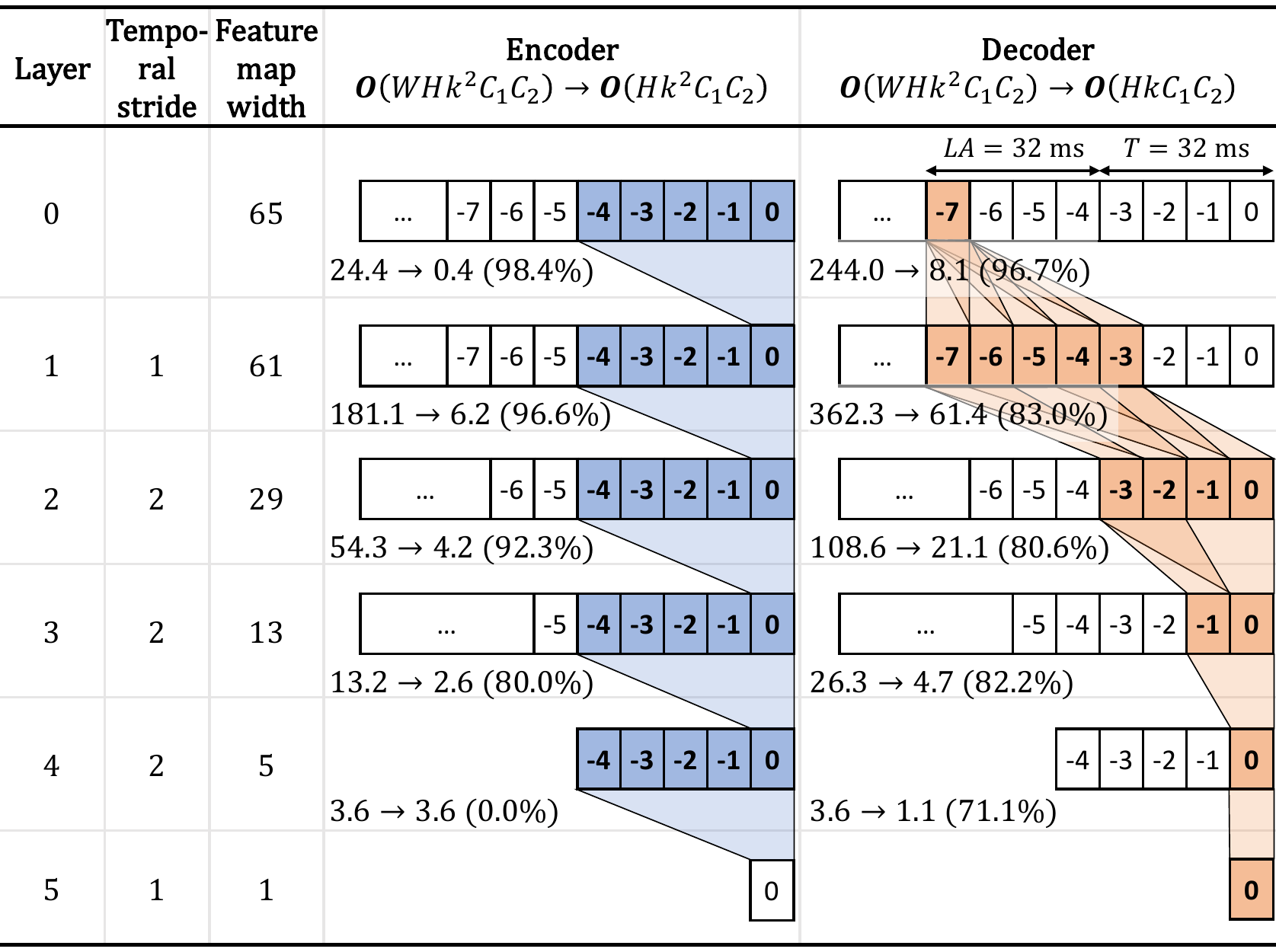}
\caption{A graphical illustration of U-Net optimization for real-time inference. As a schematic view of 2D feature map, the number in the box indicates the relative index to the latest frame. $LA$ and $T$ denote the lookahead and the frame length respectively. The number of multiplications reduced from the naïve one is shown at the bottom of the box (in millions). The overall reduction reached 88.9\%.}
\label{fig:optimization}
\vspace*{-15pt}
\end{figure}

Most computation of the naïve U-Net is concentrated on a few decoder layers before the output. Fortunately, only a single frame of the output mask is needed for real-time inference. 
Although using the latest frame can achieve the shortest latency, it is better to preview a few milliseconds for performance.
It is computationally less efficient to use a longer lookahead because more frames should be calculated in the previous decoder layer. Our real-time implementation previews 32ms which is shorter than allowed in the DNS challenge \cite{reddy2020interspeech}. The schematic details are shown in Fig. \ref{fig:optimization}.

\section{Experiments}
\label{section:experiments}
\subsection{Dataset}
We used the DNS challenge dataset \cite{reddy2020interspeech} and internally collected dataset for training.
The former is a large scale dataset where the speech samples were collected from Librivox \cite{mcguirelibrivox}, and noise samples from Audioset \cite{gemmeke2017audio} and Freesound \cite{fonseca2017freesound}. Note that we did not use the provided noisy speech data from the DNS dataset but used on-the-fly augmentation with the clean speech and noise in the two datasets during the training phase. 
Since our goal is to perform both denoising and dereverberation, we used pyroomacoustics \cite{scheibler2018pyroomacoustics} to simulate an artificial reverberation with randomly sampled absorption, room size, location of source and microphone distance.
We also trimmed random segments of 2 seconds from speech and noise data, and mixed them with uniformly distributed source-to-noise ratio (SNR) ranging from -10 dB to 30 dB.

For test, we used two datasets such as the synthesized testset in the DNS challenge (DNS) and WHAMR \cite{maciejewski2019whamr}.
The DNS synthesized testset provides noisy-reverberant mixtures and noisy mixtures without reverb.
DNS was used only to test the denoising performance since it does not provide the direct source signal of synthesized mixture samples. 
Therefore, a reverberant source was given as ground truth when the model is tested on noisy-reverberant mixtures.
Both the denoising and dereverberation performance were tested on the \textit{min} subset of WHAMR dataset which contains 3,000 audio files.
To test the denoising and dereverberation performances both simultaneously and separately, we tested our models on four scenarios: 
1) \textit{nr2d}: noisy-reverberant mixture to direct source 2) \textit{nr2r}: noisy-reverberant mixture to reverberant source 3) \textit{n2d}: noisy mixture to direct source 4) \textit{r2d}: reverberant source to direct source.
The corresponding four pair of test subsets, denoted in a following way ($\texttt{mixture}$, $\texttt{ground\_truth}$), were used as follows, 1. \textit{nr2d}: ($\texttt{mix\_single\_reverb}$, $\texttt{s1\_anechoic}$),
2. \textit{nr2r}: ($\texttt{mix\_single\_reverb}$, $\texttt{s1\_reverb}$), 3. \textit{n2d}: ($\texttt{mix\_single\_anechoic}$, $\texttt{s1\_anechoic}$), 4. \textit{r2d}: ($\texttt{s1\_reverb}$, $\texttt{s1\_anechoic}$).

\subsection{Implementation}
Input features were used as a channel-wise concatenation of log-magnitude spectrogram, real and imaginary part of demodulated phase \cite{takahashi2018phasenet}, group delay, and delta-phase \cite{mccowan2011delta}.
The window size of \texttt{model20} was 1024 with 256 hop size and the window size of \texttt{model10} was 512 with 128 hop size.
All models were trained for 125k iterations with AdamW optimizer \cite{reddi2019convergence}. The learning rate was set to 0.0004 and halved at 62.5k iteration.
Every test was done with a non-causal inference using \texttt{model20} except the experiments in subsection \ref{subsection:real-time}.

\subsection{Ablation studies}
\vspace{-10pt}
\begin{table}[ht]
\caption{The effect of proposed loss function. The \textbf{denoising} performance was tested on the DNS challenge synthesized testset (\textit{w/o} and \textit{w/} reverb) and both \textbf{denoising} and \textbf{dereverberation} performance was tested on WHAMR dataset (\textit{nr2d}: noisy-reverberant mixture to direct source, \textit{r2d}: reverberant source to direct source).}
\begin{center}
\setlength\tabcolsep{1.5pt}
\begin{tabular}{l|cc|cc|cc|cc}
\toprule
\multicolumn{9}{c}{DNS-challenge}\\
\midrule
\multicolumn{1}{c}{Loss}
&\multicolumn{2}{|c}{$\mathbb{C}$MSE}&\multicolumn{2}{|c}{SingleScale} &\multicolumn{2}{|c}{MultiScale} &\multicolumn{2}{|c}{MultiScale+}
\\ \midrule
\multicolumn{1}{c}{Reverb} &\multicolumn{1}{|c}{\textit{w/o}} &\multicolumn{1}{c}{\textit{w/}} &\multicolumn{1}{|c}{\textit{w/o}}&\multicolumn{1}{c}{\textit{w/}} &\multicolumn{1}{|c}{\textit{w/o}}&\multicolumn{1}{c}{\textit{w/}} &\multicolumn{1}{|c}{\textit{w/o} }&\multicolumn{1}{c}{\textit{w/}}
\\ \midrule
\bf{Si-SDR} & 15.63 & 14.21 & 17.47 & 15.79 & 17.57 & 15.93 & \bf{17.91} & \bf{16.22} \\
\bf{PESQ} & 2.22 & 2.59 & 2.57 & 2.90 & 2.63 & 2.97 & \bf{2.71} & \bf{3.01} \\
\toprule
\multicolumn{9}{c}{WHAMR}\\
\midrule
\multicolumn{1}{c}{Loss}
&\multicolumn{2}{|c}{$\mathbb{C}$MSE}&\multicolumn{2}{|c}{SingleScale} &\multicolumn{2}{|c}{MultiScale} &\multicolumn{2}{|c}{MultiScale+}
\\ \midrule
\multicolumn{1}{c}{Task} &\multicolumn{1}{|c}{\textit{nr2d}} &\multicolumn{1}{c}{\textit{r2d}} &\multicolumn{1}{|c}{\textit{nr2d}}&\multicolumn{1}{c}{\textit{r2d}} &\multicolumn{1}{|c}{\textit{nr2d}}&\multicolumn{1}{c}{\textit{r2d}} &\multicolumn{1}{|c}{\textit{nr2d} }&\multicolumn{1}{c}{\textit{r2d}}
\\ \midrule
\bf{Si-SDR} & 4.21 & 8.87 & 5.08 & 9.88 & 5.24 & 10.13 & \bf{5.33} & \bf{10.40} \\
\bf{PESQ} & 1.38 & 2.58 & 1.45 & 2.96 & \bf{1.54} & 3.09 & 1.52 & \bf{3.16} \\
\bottomrule
\end{tabular}
\end{center}
\label{table:loss_comparison}
\end{table}
\vspace*{-15pt}
To show the effect of loss functions we observed SI-SDR \cite{le2019sdr} and PESQ \cite{rix2001perceptual} while changing four different loss functions. Complex MSE ($\mathbb{C}$MSE) and three different cosine similarity based loss functions --- SingleScale, MultiScale, and MultiScale+ --- were compared each of which corresponds to Eq. \ref{eq:cossim}, Eq. \ref{eq:multiscale}, and Eq. \ref{eq:emphasized}, respectively.
The quantitative results in Table \ref{table:loss_comparison} show that the proposed multi-scale and emphasis functions are beneficial for both denoising and dereverberation tasks in most of the cases.
\subsection{Analysis on phase enhancement}
\begin{table}[ht]
\vspace*{-10pt}
\caption{
Phase distance and gain under four different tasks.
}
\begin{center}
\begin{tabular}{lcccc}
\toprule
Task & \textit{nr2r} & \textit{n2d} & \textit{nr2d} & \textit{r2d} \\
\midrule
$PD$($\bm{Y}$, $\bm{X}$) & 21.1\textdegree & 23.3\textdegree & 36.3\textdegree & 24.7\textdegree\\
$PD$($\bm{Y}$, $\hat{\bm{Y}}$) & 20.2\textdegree & 21.9\textdegree & 29.5\textdegree & 15.0\textdegree\\
$PhaseGain$ & 4.5\% &  6\% & 17.6\% & 64\% \\
\bottomrule
\end{tabular}
\end{center}
\label{table:phasedist}
\vspace*{-25pt}
\end{table}
\begin{figure}[ht]
\centering
\includegraphics[scale=0.4]{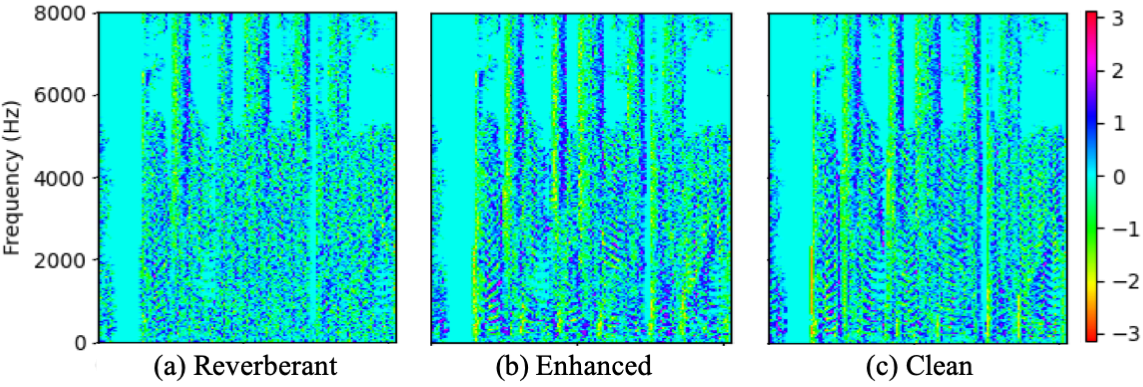}
\caption{Group delay of enhanced phase}
\label{fig:group_delay}
\vspace*{-15pt}
\end{figure}
Here, we used the phase distance defined in \cite{choi2019phase} to quantitatively measure the phase enhancement performance.
Phase distance ($PD$) between spectrogram $A$ and $B$ is formulated as follows,
\begin{equation}
\label{eq:phase_distance}
  \begin{aligned}
	PD(\bm{A}, \bm{B}) = \sum_{t,f} \frac{\abs{A_{t,f}}}{\sum_{t',f'}\abs{A_{t',f'}}} \angle(A_{t,f}, B_{t,f}),
  \end{aligned}
\end{equation}
where $\angle(A_{t,f}, B_{t,f})$ is the angle between $A_{t,f}$ and $B_{t,f}$, ranging from 0\textdegree  to 180\textdegree.
We measured $PD$ between ground truth $\bm{Y}$ and mixture $\bm{X}$, and $PD$ between ground truth $\bm{Y}$ and estimation $\bm{\hat{Y}}$, and checked how much $PhaseGain(\%)$ was obtained. This was tested on all four scenarios of WHAMR testset and shown in Table \ref{table:phasedist}.
We found that the network is able to give a reasonable $PhaseGain$ in tasks including dereverberation (\textit{nr2d}, \textit{nr2r}).
However, $PhaseGain$ was marginal for only-denoising-tasks (\textit{nr2r}, \textit{n2d}).
We conjecture that this is because the network is not able to estimate a precise magnitude value for noisy mixture and this issue is left for futurework.
A visualization of enhanced phase group delay tested on a reverberant source is shown in Fig. \ref{fig:group_delay}.
Fig. \ref{fig:group_delay} (b) shows the enhanced harmonic structure of phase group delay.

\subsection{Computation of real-time U-Net}
\label{subsection:real-time}
\vspace*{-10pt}
\begin{table}[ht]
\caption{
The effect of causality and the model size.
}
\begin{center}
\begin{tabular}{lcccc}
\toprule
Causal/Model & \xmark/NRT & \cmark/NRT & \xmark/RT & \cmark/RT \\
\midrule
\textbf{SI-SDR} & \bf{5.33} & 4.60 & 3.42 & 2.33 \\
\textbf{PESQ} & \bf{1.52} & 1.43 & 1.39 & 1.34 \\
\bottomrule
\end{tabular}
\end{center}
\label{table:real-time}
\end{table}
\vspace*{-15pt}
Following the constraint for real-time model suggested by the DNS challenge, we measured the elapsed time to compute a single frame. \texttt{model20} that took 40 ms to compute a frame will be denoted as non-real-time (NRT) model, and \texttt{model10} that took 4.32 ms to compute a frame will be denoted as real-time (RT) model. 
To compare the two models and how causal inference affects the model performance, we compare four combinations in \textit{nr2d} task. Table \ref{table:real-time} shows that both non-causal inference and model size are significant factors for performance.

Finally, we report the Mean Opinion Score (MOS) results from the DNS challenge based on the online subjective evaluation framework ITU-T P.808 \cite{p808}. 
For better perceptual quality, we linearly added the estimated direct source and reverberant source with a 15 dB ratio, and implemented a simple and zero-delay dynamic range compression to apply on it.
Our causal-NRT and causal-RT model achieved a mean opinion score of 3.36 and 3.24, respectively.

\section{Conclusions}
We proposed a new mask and loss function to improve the performance of single-stage denoising and dereverberation.
As the proposed PHM and loss function are orthogonal to the network structure, we believe that a better performance can be achieved using the variant of U-Net architectures such as \cite{takahashi2017multi, tolooshams2020channel}.



\bibliographystyle{IEEEtran}

\bibliography{mybib}

\end{document}